\documentclass[preprint]{aastex62}
\usepackage{booktabs}
\usepackage{graphicx}
\usepackage{blindtext}
\usepackage[english]{babel} 
\usepackage{amsmath,amsfonts,amsthm,bm} 
\usepackage{lineno}

\newcommand{\Spitzer}{{\it Spitzer}}

\shorttitle{KMT-2019-BLG-0371 and the Limits of Bayesian Analysis}
\shortauthors{Kim et al.}

\begin{document}

\title{KMT-2019-BLG-0371 and the Limits of Bayesian Analysis}

\correspondingauthor{Sun-Ju Chung}
\email{sjchung@kasi.re.kr, sherlock@kasi.re.kr}

\author{Yun Hak Kim}
\affiliation{Korea Astronomy and Space Science Institute, 776 Daedeokdae-ro, Yuseong-Gu, Daejeon 34055, Korea} 
\affiliation{University of Science and Technology, Korea, (UST), 217 Gajeong-ro, Yuseong-gu, Daejeon 34113, Korea}

\author{Sun-Ju Chung}
\affiliation{Korea Astronomy and Space Science Institute, 776 Daedeokdae-ro, Yuseong-Gu, Daejeon 34055, Korea}
\affiliation{University of Science and Technology, Korea, (UST), 217 Gajeong-ro, Yuseong-gu, Daejeon 34113, Korea}

\author{Jennifer C. Yee}
\affiliation{Center for Astrophysics $|$ Harvard \& Smithsonian, 60 Garden St., Cambridge, MA 02138, USA}

\author{A. Udalski  }
\affiliation{Astronomical Observatory, University of Warsaw, AI.~Ujazdowskie~4, 00-478~Warszawa, Poland}

\author{Ian A. Bond  }
\affiliation{Institute of Natural and Mathematical Science, Massey University, Auckland 0745, New Zealand} 

\author{Youn Kil Jung  }
\affiliation{Korea Astronomy and Space Science Institute, 776 Daedeokdae-ro, Yuseong-Gu, Daejeon 34055, Korea} 

\collaboration{and}
\noaffiliation

\author{Andrew Gould  }
\affiliation{Korea Astronomy and Space Science Institute, 776 Daedeokdae-ro, Yuseong-Gu, Daejeon 34055, Korea}
\affiliation{Department of Astronomy, Ohio State University, 140 W. 18th Ave., Columbus, OH 43210, USA}
\affiliation{Max-Planck-Institute for Astronomy, K{\"o}nigstuhl 17, 69117 Heidelberg, Germany}

\author{Michael D. Albrow  }
\affiliation{Department of Physics and Astronomy, University of Canterbury, Private Bag 4800 Christchurch, New Zealand} 

\author{Cheongho Han  }
\affiliation{Department of Physics, Chungbuk National University, Cheongju 28644, Korea}

\author{Kyu-Ha Hwang  }
\affiliation{Korea Astronomy and Space Science Institute, 776 Daedeokdae-ro, Yuseong-Gu, Daejeon 34055, Korea}

\author{Yoon-Hyun Ryu  }
\affiliation{Korea Astronomy and Space Science Institute, 776 Daedeokdae-ro, Yuseong-Gu, Daejeon 34055, Korea}

\author{In-Gu Shin  }
\affiliation{Korea Astronomy and Space Science Institute, 776 Daedeokdae-ro, Yuseong-Gu, Daejeon 34055, Korea}

\author{Yossi Shvartzvald }
\affiliation{Department of Particle Physics and Astrophysics, Weizmann Institute of Science, Rehovot 76100, Israel}

\author{Weicheng Zang  }
\affiliation{Department of Astronomy and Tsinghua Centre for Astrophysics, Tsinghua University, Beijing 100084, China}

\author{Sang-Mok Cha  }
\affiliation{Korea Astronomy and Space Science Institute, 776 Daedeokdae-ro, Yuseong-Gu, Daejeon 34055, Korea}
\affiliation{School of Space Research, Kyung Hee University, Giheung-gu, Yongin, Gyeonggi-do, 17104, Korea}

\author{Dong-Jin Kim  }
\affiliation{Korea Astronomy and Space Science Institute, 776 Daedeokdae-ro, Yuseong-Gu, Daejeon 34055, Korea}

\author{Hyoun-Woo Kim  }
\affiliation{Korea Astronomy and Space Science Institute, 776 Daedeokdae-ro, Yuseong-Gu, Daejeon 34055, Korea}
\affiliation{Department of Astronomy and Space Science, Chungbuk National University, Cheongju 28644, Republic of Korea}

\author{Seung-Lee Kim  }
\affiliation{Korea Astronomy and Space Science Institute, 776 Daedeokdae-ro, Yuseong-Gu, Daejeon 34055, Korea} 
\affiliation{University of Science and Technology, Korea, (UST), 217 Gajeong-ro, Yuseong-gu, Daejeon 34113, Korea}

\author{Chung-Uk Lee  }
\affiliation{Korea Astronomy and Space Science Institute, 776 Daedeokdae-ro, Yuseong-Gu, Daejeon 34055, Korea} 

\author{Dong-Joo Lee  }
\affiliation{Korea Astronomy and Space Science Institute, 776 Daedeokdae-ro, Yuseong-Gu, Daejeon 34055, Korea} 

\author{Yongseok Lee  }
\affiliation{Korea Astronomy and Space Science Institute, 776 Daedeokdae-ro, Yuseong-Gu, Daejeon 34055, Korea}
\affiliation{School of Space Research, Kyung Hee University, Giheung-gu, Yongin, Gyeonggi-do, 17104, Korea}

\author{Byeong-Gon Park  }
\affiliation{Korea Astronomy and Space Science Institute, 776 Daedeokdae-ro, Yuseong-Gu, Daejeon 34055, Korea} 
\affiliation{University of Science and Technology, Korea, (UST), 217 Gajeong-ro, Yuseong-gu, Daejeon 34113, Korea}

\author{Richard W. Pogge  }
\affiliation{Department of Astronomy, Ohio State University, 140 W. 18th Ave., Columbus, OH 43210, USA}

\collaboration{(KMTNet Collaboration),}
\noaffiliation

\author{Radek Poleski  }
\affiliation{Astronomical Observatory, University of Warsaw, AI.~Ujazdowskie~4, 00-478~Warszawa, Poland}

\author{Przemek Mr{\'o}z  }
\affiliation{Division of physics, Mathematics, and Astronomy, California institute of Technology, Pasadena, CA 91125, USA}

\author{Jan Skowron  }
\affiliation{Astronomical Observatory, University of Warsaw, AI.~Ujazdowskie~4, 00-478~Warszawa, Poland}

\author{Michal K. Szyma{\'n}ski  }
\affiliation{Astronomical Observatory, University of Warsaw, AI.~Ujazdowskie~4, 00-478~Warszawa, Poland}

\author{Igor Soszy{\'n}ski  }
\affiliation{Astronomical Observatory, University of Warsaw, AI.~Ujazdowskie~4, 00-478~Warszawa, Poland}

\author{Pawel Pietrukowicz  }
\affiliation{Astronomical Observatory, University of Warsaw, AI.~Ujazdowskie~4, 00-478~Warszawa, Poland}

\author{Szymon Koz{\l}owski  }
\affiliation{Astronomical Observatory, University of Warsaw, AI.~Ujazdowskie~4, 00-478~Warszawa, Poland}

\author{Krzysztof Ulaczyk  }
\affiliation{Astronomical Observatory, University of Warsaw, AI.~Ujazdowskie~4, 00-478~Warszawa, Poland}
\affiliation{Department of Physics, University of Warwick, Gibbet Hill Road, Coventry, CV4~7AL,~UK}

\author{Krzysztof A. Rybicki  }
\affiliation{Astronomical Observatory, University of Warsaw, AI.~Ujazdowskie~4, 00-478~Warszawa, Poland}

\author{Patryk Iwanek  }
\affiliation{Astronomical Observatory, University of Warsaw, AI.~Ujazdowskie~4, 00-478~Warszawa, Poland}

\author{Marcin Wrona }
\affiliation{Astronomical Observatory, University of Warsaw, AI.~Ujazdowskie~4, 00-478~Warszawa, Poland}

\author{Mariusz Gromadzki }
\affiliation{Astronomical Observatory, University of Warsaw, AI.~Ujazdowskie~4, 00-478~Warszawa, Poland}

\collaboration{(OGLE Collaboration),}
\noaffiliation

\author{Fumio Abe  }
\affiliation{Institute for Space-Earth Environmental Research, Nagoya University, Nagoya 464-8601, Japan}

\author{Richard Barry  }
\affiliation{Code 667, NASA Goddard Space Flight Center, Greenbelt, MD 20771, USA}

\author{David P. Bennett  }
\affiliation{Code 667, NASA Goddard Space Flight Center, Greenbelt, MD 20771, USA}
\affiliation{Department of Astronomy, University of Maryland, College Park, MD 20742, USA}

\author{Aparna Bhattacharya }
\affiliation{Code 667, NASA Goddard Space Flight Center, Greenbelt, MD 20771, USA}
\affiliation{Department of Astronomy, University of Maryland, College Park, MD 20742, USA}

\author{Martin Donachie  }
\affiliation{Department of Physics, University of Auckland, Private Bag 92019, Auckland, New Zealand}

\author{Hirosane Fujii  }
\affiliation{Department of Earth and Space Science, Graduate School of Science, Osaka University, Toyonaka, Osaka 560-0043, Japan}

\author{Akihiko Fukui  }
\affiliation{Department of Earth and Planetary Science, Graduate School of Science, The University of Tokyo, 7-3-1 Hongo, Bunkyo-ku, Tokyo 113-0033, Japan}
\affiliation{Instituto de Astrof\'isica de Canarias, V\'ia L\'actea s/n, E-38205 La Laguna, Tenerife, Spain}

\author{Yoshitaka Itow  }
\affiliation{Institute for Space-Earth Environmental Research, Nagoya University, Nagoya 464-8601, Japan}

\author{Yuki Hirao  }
\affiliation{Department of Earth and Space Science, Graduate School of Science, Osaka University, Toyonaka, Osaka 560-0043, Japan}

\author{Rintaro Kirikawa  }
\affiliation{Department of Earth and Space Science, Graduate School of Science, Osaka University, Toyonaka, Osaka 560-0043, Japan}

\author{Iona Kondo  }
\affiliation{Department of Earth and Space Science, Graduate School of Science, Osaka University, Toyonaka, Osaka 560-0043, Japan}

\author{Naoki Koshimoto  }
\affiliation{Department of Astronomy, Graduate School of Science, The University of Tokyo, 7-3-1 Hongo, Bunkyo-ku, Tokyo 113-0033, Japan}
\affiliation{National Astronomical Observatory of Japan, 2-21-1 Osawa, Mitaka, Tokyo 181-8588, Japan}

\author{Yutaka Matsubara  }
\affiliation{Institute for Space-Earth Environmental Research, Nagoya University, Nagoya 464-8601, Japan}

\author{Yasushi Muraki  }
\affiliation{Institute for Space-Earth Environmental Research, Nagoya University, Nagoya 464-8601, Japan}

\author{Shota Miyazaki  }
\affiliation{Department of Earth and Space Science, Graduate School of Science, Osaka University, Toyonaka, Osaka 560-0043, Japan}

\author{Cl{\'e}ment Ranc  }
\affiliation{Code 667, NASA Goddard Space Flight Center, Greenbelt, MD 20771, USA}

\author{Nicholas J. Rattenbury  }
\affiliation{Department of Physics, University of Auckland, Private Bag 92019, Auckland, New Zealand}

\author{Yuki Satoh  }
\affiliation{Department of Earth and Space Science, Graduate School of Science, Osaka University, Toyonaka, Osaka 560-0043, Japan}

\author{Hikaru Shoji  }
\affiliation{Department of Earth and Space Science, Graduate School of Science, Osaka University, Toyonaka, Osaka 560-0043, Japan}

\author{Takahiro Sumi  }
\affiliation{Department of Earth and Space Science, Graduate School of Science, Osaka University, Toyonaka, Osaka 560-0043, Japan}

\author{Daisuke Suzuki  }
\affiliation{Department of Earth and Space Science, Graduate School of Science, Osaka University, Toyonaka, Osaka 560-0043, Japan}

\author{Paul J. Tristram  }
\affiliation{University of Canterbury Mt. John Observatory, P.O.Box 56, Lake Tekapo 8770, New Zealand}

\author{Yuzuru Tanaka  }
\affiliation{Department of Earth and Space Science, Graduate School of Science, Osaka University, Toyonaka, Osaka 560-0043, Japan}

\author{Tsubasa Yamawaki  }
\affiliation{Department of Earth and Space Science, Graduate School of Science, Osaka University, Toyonaka, Osaka 560-0043, Japan}

\author{Atsunori Yonehara  }
\affiliation{Department of Physics, Faculty of Science, Kyoto Sangyo University, Kyoto 603-8555, Japan}

\collaboration{(MOA Collaboration)}
\noaffiliation

\begin{abstract}
We show that the perturbation at the peak of the light curve of microlensing event KMT-2019-BLG-0371 is explained by a model with a mass ratio between the host star and planet of $q \sim 0.08$. Due to the short event duration ($t_{\rm E} \sim 6.5\ $ days), the secondary object in this system could potentially be a massive giant planet. A Bayesian analysis shows that the system most likely consists of a host star with a mass $M_{\rm h} = 0.09^{+0.14}_{-0.05}M_{\odot}$ and a massive giant planet with a mass $M_{\rm p} = 7.70^{+11.34}_{-3.90}M_{\rm Jup}$. However, the interpretation of the secondary as a planet (i.e., as having $M_{\rm p} < 13 M_{\rm Jup}$) rests entirely on the Bayesian analysis. Motivated by this event, we conduct an investigation to determine which constraints meaningfully affect Bayesian analyses for microlensing events. We find that the masses inferred from such a Bayesian analysis are determined almost entirely by the measured value of $\theta_{\rm E}$ and are relatively insensitive to other factors such as the direction of the event $(\ell, b)$, the lens-source relative proper motion $\mu_{\rm rel}$, or the specific Galactic model prior.
\end{abstract}


\bigskip
\section{Introduction}
In microlensing events, it is rare to measure the flux from the lens star; even if it is detected in normal, seeing-limited data, it is rare to be able to definitively identify it with the lens (however, see e.g., Han et al\citealt{han2019a}). Without a measurement of the lens flux, it is impossible to infer its mass through standard techniques, e.g., by putting the lens on a color-magnitude diagram. However, there are a few cases in which both the microlens parallax, $\pi_{\rm E}$, and the finite source effect (leading to a measurement of the Einstein ring radius, $\theta_{\rm E}$) may be measured (Refsdal\citealt{refsdal1966}; Gould\citealt{gould1992}, Gould\citealt{gould1994}; Wiit $\&$ Mao\citealt{wittmao1994}; Nemiroff $\&$ Wickramasinghe\citealt{NW1994}; Gould\citealt{gould2000}). In such cases, the lens mass may be computed as

\begin{equation}
M = \frac{\theta_{\rm E}}{\kappa\pi_{\rm E}},
\label{eqn:one}
\end{equation}
where $\kappa = 8.14\ \mathrm{mas}\ M_{\odot}^{-1}$. 

At the same time, the microlens parallax often cannot be measured. As a result, the mass of the lens is unmeasured. If there is a planetary perturbation in the event, the planet's mass is measured only relative to the mass of the lens star, and hence, the mass of the planet is also unmeasured. If we use a mass cut, e.g., $13 M_{\rm Jup}$, to define the boundary between a planet and a brown dwarf, then this ambiguity can lead to ambiguity as to whether a given companion is a planet or not.

In this paper, we present the analysis of one such event. The mass ratio, $q$, between the companion and the primary in KMT-2019-BLG-0371 is $q \sim 0.08$. Hence, the companion will be of ``planetary" mass only if the primary is $M_{\rm L} \lesssim 0.1 M_{\odot}$. However, because the microlens parallax is not measured, we must use a Bayesian analysis to estimate the mass of the lens, as is standard practice. 

Because our interpretation of this companion rests on the Bayesian analysis, we use this event to study how various inputs to such an analysis affect the inferred mass. This investigation is the first to systematically investigate the role of the physical constraints from the lens-source relative proper motion, $\mu_{\rm rel}$, and $\theta_{\rm E}$ in this kind of Bayesian inference and stands to provide insight into the method as a whole. Nevertheless, because this investigation is motivated by KMT-2019-BLG-0371, we first begin by describing the analysis of that event.

\bigskip
\section{Observations}
The microlensing event KMT-2019-BLG-0371 was first discovered on April 4 by the alert system of the Korea Microlensing Telescope Network (KMTNet; Kim et al.\citealt{kim2016}, Kim et al.\citealt{kim2018}). KMTNet uses three identical ground-based telescopes located at the Cerro Tololo Inter-American Observatory in Chile (KMTC), South African Astronomical Observatory in South Africa (KMTS), and Siding Spring Observatory in Austrailia (KMTA), respectively. The event was detected in two KMTNet fields (BLG01 and BLG41), which overlap each other and have a combined cadence of $\Gamma =4$ hr$^{-1}$. The event was mainly observed with {\it{I}} band images, and some {\it{V}} band images were taken as well for the color determination of the source star. The KMTNet data were reduced by pySIS (Albrow et al.\citealt{albrow2009}), which is based on the Difference Image Analysis (DIA; Alard \& Lupton\citealt{alard1998}, Alard\citealt{alard2000}).

The event was also detected by the Optical Gravitational Lensing Experiment (OGLE; Udalski\citealt{udalski2003}; Udalski et al.\citealt{udalski2015}) by 1.3m Warsaw telescope at the Las Campanas Observatory in Chile and is designated as OGLE-2019-BLG-0505. The equatorial positions of the event are $\rm{(R.A.,decl)}_{\rm{J2000}}=(17^{\rm{h}}53^{\rm{m}}32^{\rm{s}}.14,-31\degr 33\arcmin 18\farcs 08)$ and the corresponding galactic coordinates are $(l, b)=(358\fdg 630,-2\fdg 811).$ 

Moreover, the event was found by the Microlensing Observations in Astrophysics (MOA; Sumi et al.\citealt{sumi2016}), and it was listed as MOA-2019-BLG-168. Hence, three different data sets were used to analyze the event. 

The three different data sets were produced from the photometry pipeline of each observatory. Thus, the data need the re-normalization process described in Yee et al.\citet{yee2012}, which is done by

\begin{equation}
\sigma'=k\sqrt{\sigma_{\it i}^2+(\sigma_{0})^2},
\label{eqn:yee}
\end{equation}
where $\sigma_{\it i}$ is the original error of the data, {\it k} and $\sigma_{0}$ is the re-normalizing parameters in order to achieve $\chi^{2}/$dof$\ =1$. The re-normalizing parameters for each observatory are introduced in Table \ref{table:obs}.\\

\begin{deluxetable}{lccc}
\tablecolumns{4} \tablewidth{0pt} \tablecaption{\textsc{Data and error re-normalization parameters}}
\tablehead{\colhead{Observatory (Band)} & \colhead{$N_{\rm{data}}$} & \colhead{\it{k}} & \colhead{$\sigma_{0}\ (\rm{mag})$}}
\startdata
OGLE       ({\rm I}) & 2254 & 1.597 & 0.003 \\
KMTC BLG01 ({\rm I}) & 2234 & 1.411 & 0.000 \\
KMTC BLG41 ({\rm I}) & 2189 & 1.381 & 0.000 \\
KMTA BLG01 ({\rm I}) & 1276 & 1.352 & 0.000 \\
KMTA BLG41 ({\rm I}) & 1489 & 1.569 & 0.000 \\
KMTS BLG01 ({\rm I}) & 1270 & 1.189 & 0.000 \\
KMTS BLG41 ({\rm I}) & 1520 & 1.412 & 0.000 \\
MOA        ({\rm R}) & 1871 & 1.041 & 0.000 \\
\enddata
\tablecomments{$N_{\rm{data}}$ is the number of data points.}
\label{table:obs}
\end{deluxetable}

\section{Light Curve Analysis}
\begin{figure}
\plotone{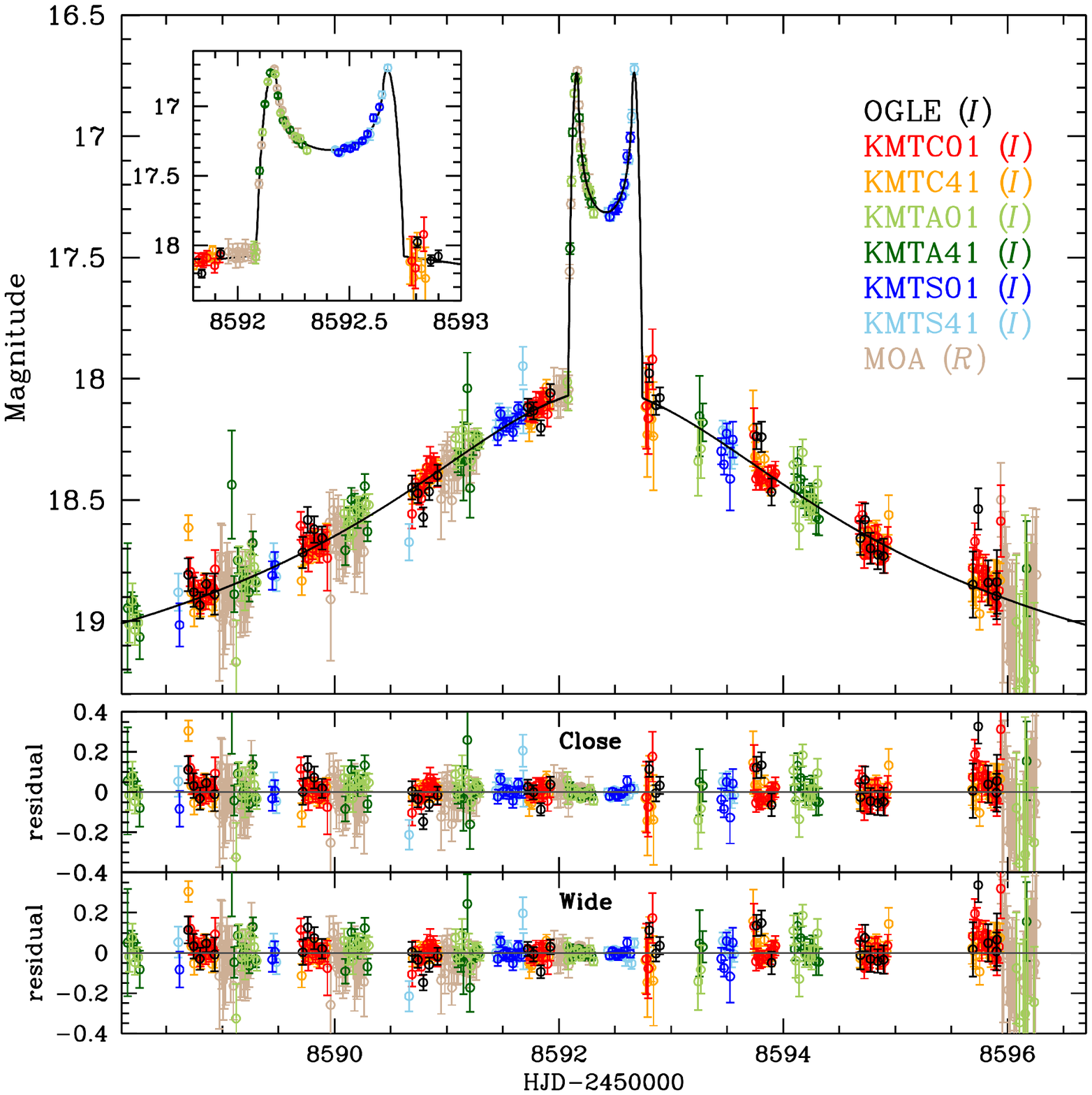}
\caption{Light curve of the best-fit binary model for KMT-2019-BLG-0371. The lower two panels show the residuals from the close and wide solutions.}
\label{fig:one}
\end{figure}

As shown in Figure \ref{fig:one}, the light curve of the microlensing event KMT-2019-BLG-0371 was well covered by three different microlensing survey groups. The caustic-crossing feature is clearly seen around HJD$'\sim8592.4$, which lasts relatively short time ($\sim0.6$ days). From this, one can infer that the lens is probably a binary system with a low-mass secondary companion. Hence, we carry out the binary lens modeling. Binary lens modeling requires seven basic parameters including single lens parameters $(t_{0},u_{0},t_{\rm E})$, binary lens parameters $(s,q,\alpha)$, and the normalized source radius $\rho = \theta_{\ast}/\theta_{\rm E}$, in which $\theta_{\ast}$ is the angular radius of the source star. Here $t_{0}$ is the time of the closest source approach to the lens, $u_{0}$ is the impact parameter in units of $\theta_{\rm E}$, $t_{\rm E}$ is the Einstein radius crossing time of the event, $s$ is the projected separation of the lens components in units of $\theta_{\rm E}$, $q$ is the mass ratio between the two lens components, and $\alpha$ is the angle between the source trajectory and the binary axis. 

\begin{figure}
\plotone{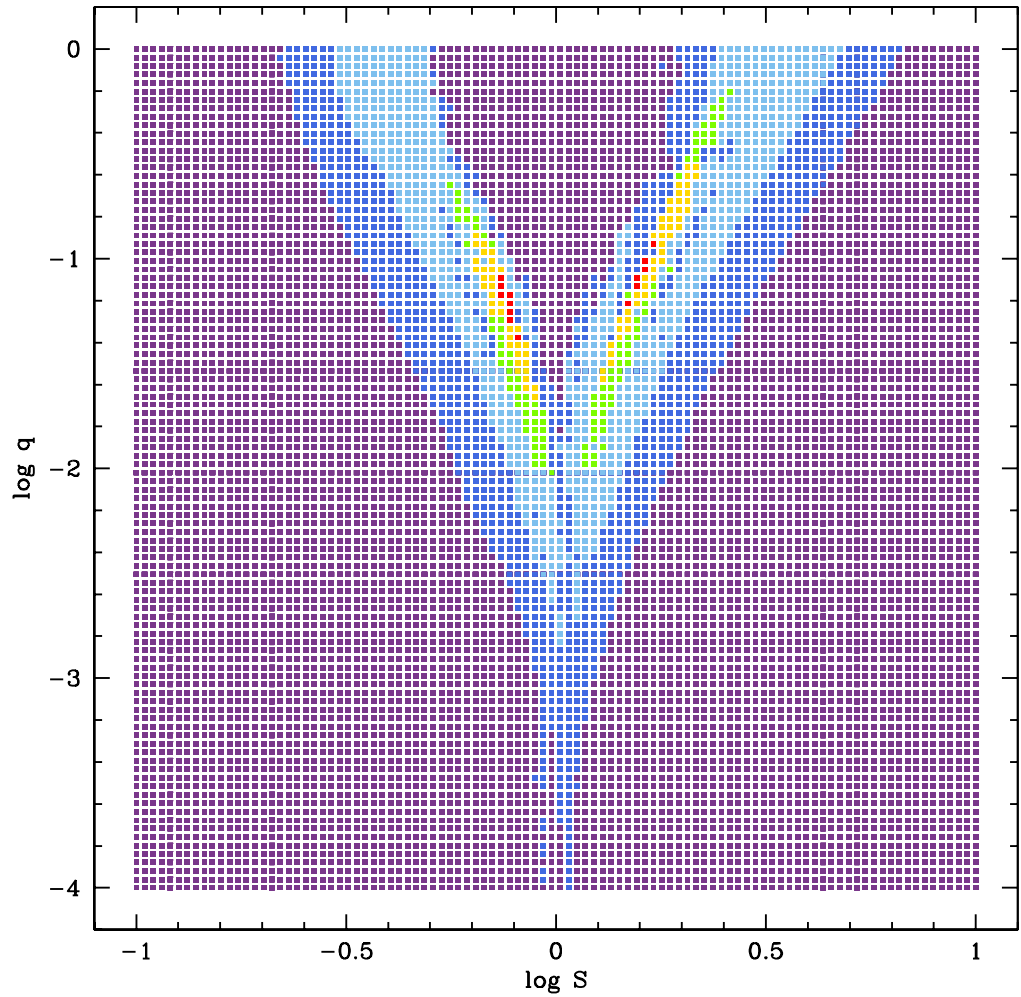}
\caption{$\Delta\chi^{2}$ distributions in the $(\log s, \log q)$ plane obtained from the grid search. The colors (red, yellow, green, sky blue, vivid blue, purple) indicate $\Delta \chi^{2} < (5^2, 10^2, 15^2, 20^2, 25^2, 30^2)$, respectively.}
\label{fig:two}
\end{figure}

We first conduct a grid search for three parameters ($s$,$q$,$\alpha$) to find local $\Delta\chi^{2}$ minima through a Markov Chain Monte Carlo (MCMC) method. The ranges of each parameter are $-1 \leq \log s \leq 1$, $-4.0 \leq \log q \leq 0$, and $0 \leq \alpha \leq 2\pi$, respectively. During the grid search, $(s,q)$ are fixed and the other parameters are allowed to vary in a MCMC chain. As a result, we find two local solutions with $(s,q)=(0.83,0.08)$ and $(s,q)=(1.58,0.12)$ (see Figure \ref{fig:two}), which are caused by a well-known close/wide degeneracy $(s\leftrightarrow 1/s)$ (Griest \& Safizadeh\citealt{griest1998}; Dominik\citealt{dominik1999}).  

\begin{figure}
\plotone{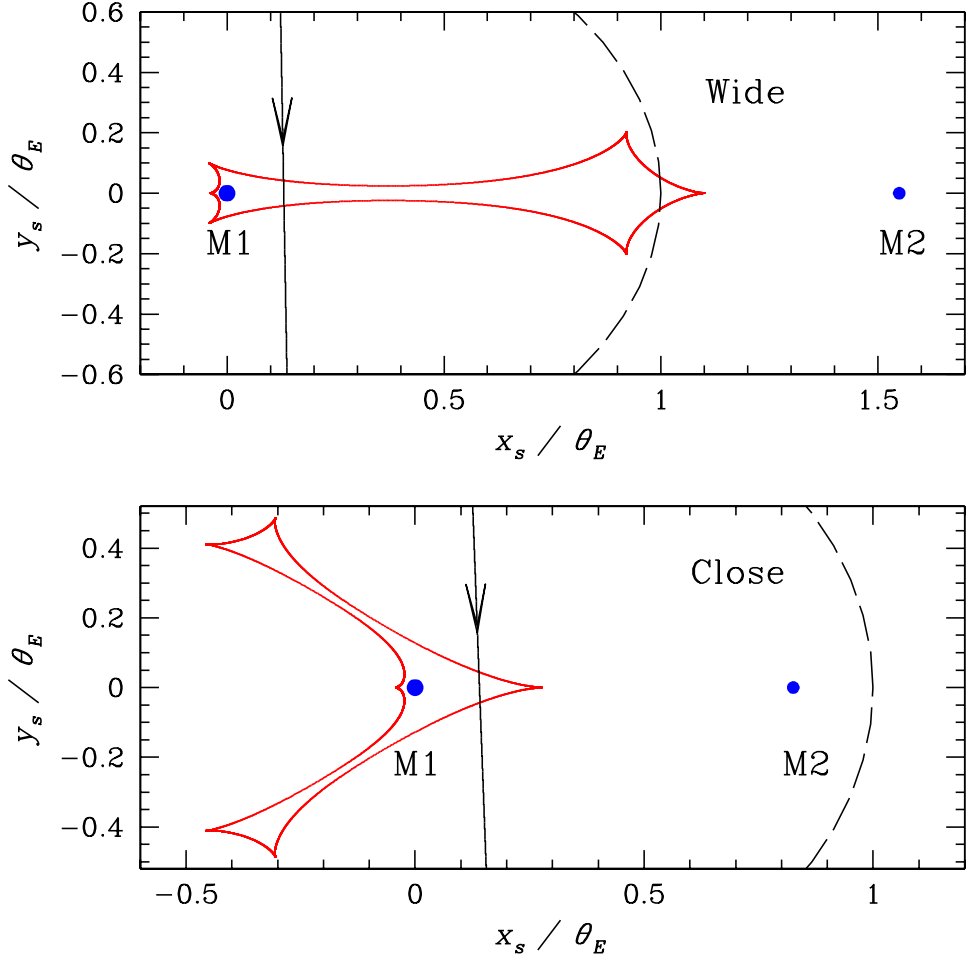}
\caption{Best-fit geometries of close/wide solutions. The red closed curves indicate the resonant caustic. $M_{1}$ and $M_{2}$ are the lens star and planetary companion.}
\label{fig:three}
\end{figure}

Based on the two local solutions, we conduct additional modeling for which all parameters are allowed to vary. The best-fit parameters for the close and wide solutions are listed in Table \ref{table:bestfit}. The difference in $\chi^{2}$ between the two solutions is less than 2.5, thus making the close/wide degeneracy severe. Figure \ref{fig:three} represents the geometries of the close and wide solutions. The source for the two solutions passes the cusps of the resonant caustic and this generates the caustic-crossing feature in the best-fit light curve (see Figure \ref{fig:one}).

We also carry out further modeling with the microlens parallax effect (Gould\citealt{gould1992}) and orbital motion effect (Dominik\citealt{dominik1998}), even though the event timescale $t_{\rm E}\sim6.5\, \rm days$ is short compared to the typical expected timescales for measuring the two effects. In order to consider these two effects in modeling, we need four additional parameters, which are defined as $\bm{\pi_{\rm{E}}}=(\pi_{\rm{E,N}}, \pi_{\rm{E,E}})$ and $(ds/dt, d\alpha/dt)$. Here, $(\pi_{\rm{E,N}},\pi_{\rm{E,E}})$ are the north and east components of the microlens parallax vector and $(ds/dt,d\alpha/dt)$ are the instantaneous change rates of $s$ and $\alpha$, respectively. As a result, there is no significant improvement $(\Delta\chi^{2} < 5$) compared to the standard model, as we expected. Hence, we perform a Bayesian analysis to estimate the physical parameters of the lens system, which will be discussed in Section 4.2.

\begin{deluxetable}{lcc}
\tablecolumns{3} \tablewidth{0pc} \tablecaption{\textsc{Best fit parameters}}
\tablehead{\colhead{Parameters} & \colhead{Close} & \colhead{Wide}} 
\startdata
$\chi^2/\rm dof$     & $13784.877/14080$         &   $13787.268/14080$    \\
$t_{0}$ (HJD$'$)     & $8592.392   \pm 0.006$    &   $8592.391 \pm 0.006$ \\
$u_{0}$              & $0.140      \pm 0.007$    &   $0.141    \pm 0.007$ \\
$t_{\rm E}$ (days)   & $6.50       \pm 0.15 $    &   $6.57     \pm 0.15 $ \\
$s$                  & $0.826      \pm 0.011$    &   $1.575    \pm 0.014$ \\
$q \ (M_{2}/M_{1})$  & $0.079      \pm 0.002$    &   $0.123    \pm 0.003$ \\
$\alpha \ (\rm rad)$ & $1.542      \pm 0.008$    &   $1.544    \pm 0.007$ \\
$\rho \ (10^{-3})$   & $6.46       \pm 0.17$     &   $6.99     \pm 0.21 $ \\
$f_{\rm{s,OGLE}}$     & $0.190      \pm 0.008$    &   $0.206    \pm 0.008$ \\
$f_{\rm{b,OGLE}}$     & $0.302      \pm 0.012$    &   $0.329    \pm 0.013$ \\
\enddata
\tablecomments{\ HJD$'$ = HJD - 2450000. $f_{\rm s,OGLE}$ and $f_{\rm b,OGLE}$ are the source and blend fluxes of OGLE data sets, in which they are determined from the modeling.}
\label{table:bestfit}
\end{deluxetable}

\bigskip
\section{Physical Properties}

\subsection{Angular Einstein Radius}
The determination of the angular Einstein radius $\theta_{\rm E}$ requires measurements of the normalized source radius $\rho$ and angular source radius $\theta_{\ast}$. Thanks to the caustic-crossing features, we measure $\rho$ from the modeling, while $\theta_{\ast}$ is determined from the intrinsic color and magnitude of the source star. For the determination of $\theta_{\ast}$, we follow the procedure of Yoo et al.\citet{yoo2004}.

\begin{figure}
\plotone{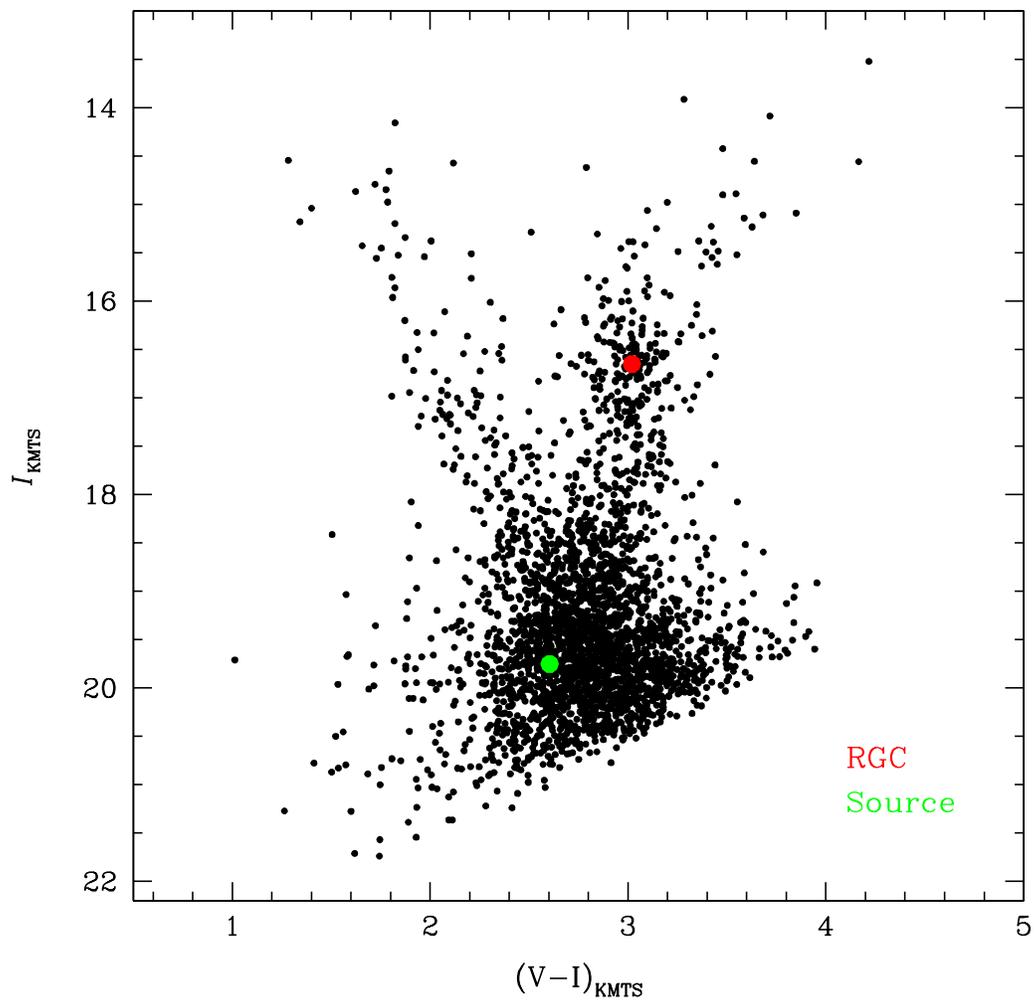}
\caption{Instrumental color-magnitude diagram (CMD) of the event KMT-2019-BLG-0371. The red and green dots represent the giant clump centroid and source star, respectively.}
\label{fig:four}
\end{figure}

First, we construct the instrumental KMTS color-magnitude diagram (CMD) using the pyDIA pipeline, which is shown in Figure \ref{fig:four}. In general, KMTC data are used for CMDs because CTIO has better seeing. However, for this event, we use KMTS data, because only KMTS has a few $V$-band data points in the caustic-crossing region. We then estimate the centroid of the giant clump from the CMD and find the instrumental color and magnitude of the source star from the best-fit model. The clump centroid is $(V-I,I)_{\rm GC} = (3.02,16.65)$, while the instrumental color and magnitude of the source are $(V-I,I)_{\rm s, close}=(2.60,19.75)$ and $(V-I,I)_{\rm s, wide}=(2.62,19.65)$ for the close and wide solutions, respectively. Then the intrinsic color and magnitude of the source are estimated as 
\begin{equation}
(V-I, I)_{\rm s,0}=(V-I, I)_{\rm s}-(V-I, I)_{\rm GC}+(V-I, I)_{\rm GC,0},
\end{equation}
where the intrinsic color and magnitude of the clump is $(V-I,I)_{\rm GC,0}=(1.06,14.52)$, which are determined from  Bensby et al.\citet{bensby2013} and Nataf et al.\citet{nataf2013}. As a result, we find that $(V-I, I)_{\rm s,0,close} = (0.64 \pm 0.07, 17.62 \pm 0.09)$ and $(V-I, I)_{\rm s,0,wide} = (0.66 \pm 0.07, 17.52 \pm 0.09)$. This indicates that the source is a late F-type or an early G-type dwarf (Bessell $\&$ Brett\citealt{bessell1988}). From the $\it{VIK}$ color-color relation (Bessell $\&$ Brett\citealt{bessell1988}) and the color-surface brightness relation (Kervella et al.\citealt{kervella2004}), we estimate the angular source radius as
\begin{eqnarray}
\theta_{\ast}&=&\left\{\begin{array}{rl} 
0.883\pm 0.078 \ \mu{\rm as} &\mbox{\rm{\ (close)}}\\
0.946\pm 0.083 \ \mu{\rm as} &\mbox{\rm{\ (wide),}}
\end{array} \right.
\end{eqnarray}
and then the angular Einstein radii for the close and wide solutions are determined as
\begin{eqnarray}
\theta_{\rm E}=\frac{\theta_{\ast}}{\rho}&=&\left\{\begin{array}{rl}
0.137\pm 0.012 \ {\rm mas} &\mbox{\rm{\ (close)}}\\ 
0.135\pm 0.013 \ {\rm mas} &\mbox{\rm{\ (wide).}}
\end{array} \right.
\end{eqnarray}
The relative lens-source proper motion is
\begin{eqnarray}
\mu_{\rm rel}= \frac{\theta_{\rm E}}{t_{\rm E}}&=&\left\{\begin{array}{rl}
7.68 \pm 0.70 \ \rm{mas} \ yr^{-1} &\mbox{\rm{\ (close)}}\\
7.53 \pm 0.70 \ \rm{mas} \ yr^{-1} &\mbox{\rm{\ (wide)}}. 
\end{array} \right.
\end{eqnarray}

\begin{deluxetable}{lcc}
\tablecolumns{3} \tablewidth{0pt} \tablecaption{\textsc{Source parameters}}
\tablehead{\colhead{Parameters}   &   \colhead{close}   &  \colhead{wide}  }
\startdata
$(V-I)_{\rm s,0}$                &  $0.64   \pm 0.07$  &  $0.66   \pm 0.07$\\
$I_{\rm s,0}$                    &  $17.62  \pm 0.09$  &  $17.52  \pm 0.09$\\
$\theta_{\ast} \ (\mu \rm as)$   &  $0.88   \pm 0.08$  &  $0.95   \pm 0.08$\\
\enddata
\label{table:source}
\end{deluxetable}

The parameters related to the source star are listed in Table 3.

\bigskip
\subsection{Lens properties}
As mentioned in Section 1, the lens mass $M$ can be measured from the measurements of the angular Einstein radius $\theta_{\rm E}$ and microlens parallax $\pi_{\rm E}$ (see Equation (\ref{eqn:one})). Using these two parameters, one can also measure the distance to the lens $D_{\rm L}$, because it is defined as

\begin{equation}
D_{\rm L}=\frac{\rm au}{\pi_{\rm E}\theta_{\rm E}+\pi_{\rm S}}, 
\label{lens1}
\end{equation}
where $\pi_{\rm S}={\rm au}/D_{\rm S}$ is the source parallax (Gould\citealt{gould2004}). Here we adopt $D_{\rm S}=8.40 \ {\rm kpc}$ from Nataf et al.\citet{nataf2013}. Although $\theta_{\rm E}$ was measured from the detection of the caustic-crossing feature, $\pi_{\rm E}$ was not measured due to the short timescale $t_{\rm E} \sim 6.5$ days. Therefore, we perform a Bayesian analysis to estimate the physical parameters of the lens. For the Bayesian analysis, we follow the procedures of Jung et al.\citet{jung2018a} using the Galactic model that is based on the velocity distribution (Han $\&$ Gould\citealt{hangould1995}), mass function (Chabrier\citealt{chabrier2003}), and matter density profile of the Galaxy (Han $\&$ Gould\citealt{hangould2003}).

\begin{figure}
\plotone{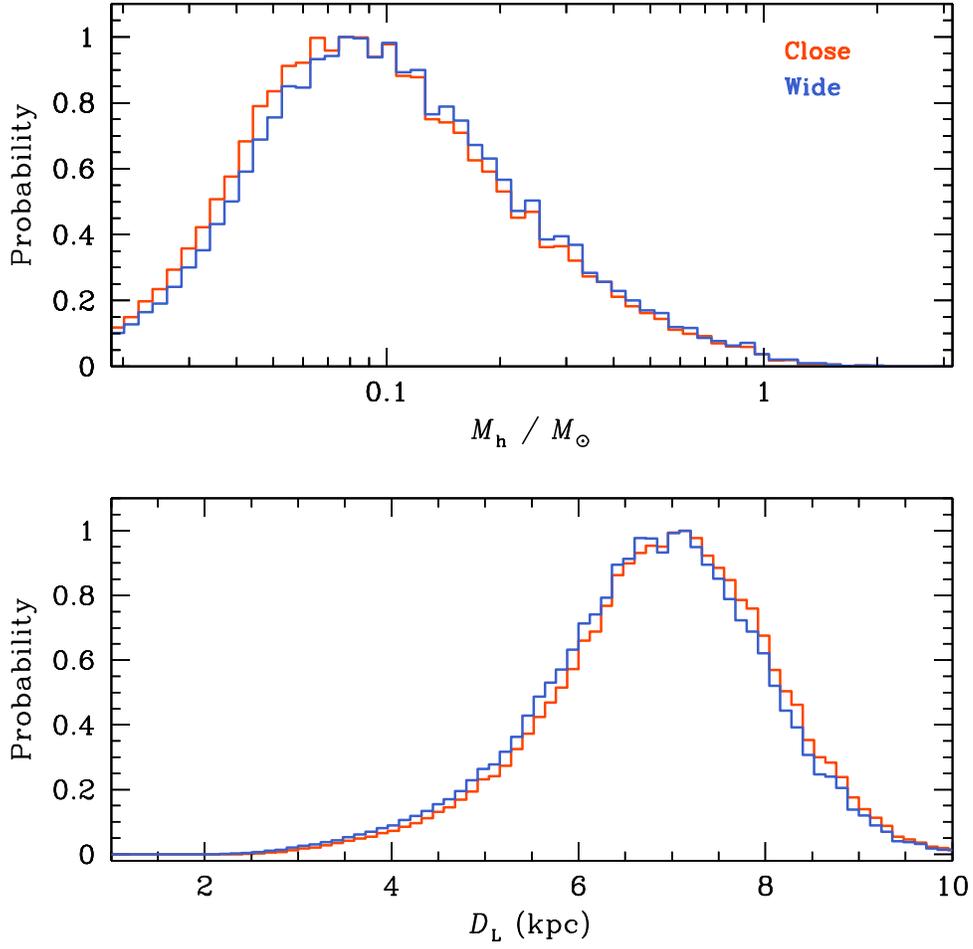}
\caption{Bayesian probability distributions of the host mass $(M_{\rm h})$ and the distance $(D_{\rm L})$. Red and Blue lines indicate the close and wide solutions, respectively.}
\label{fig:five}
\end{figure}

Figure \ref{fig:five} shows the Bayesian probability distributions for the physical parameters of the host star for close and wide solutions. The mass and distance of the host star are 

\begin{eqnarray}
M_{\rm h}&=&\left\{\begin{array}{rl}
0.093^{+0.137}_{-0.049}\ M_{\odot} &\mbox{\rm{\ (close)}}\\ 
0.093^{+0.137}_{-0.049}\ M_{\odot} &\mbox{\rm{\ (wide),}}
\end{array} \right.
\end{eqnarray}
and
\begin{eqnarray}
D_{\rm{L}}&=&\left\{\begin{array}{rl}
7.01^{+1.05}_{-1.21}\ \rm{kpc} &\mbox{\rm{\ (close)}}\\ 
7.04^{+1.05}_{-1.20}\ \rm{kpc} &\mbox{\rm{\ (wide).}}
\end{array} \right.
\end{eqnarray}
Because of $M_p = qM_{\rm h}$, the secondary companion mass of the host star is
\begin{eqnarray}
M_{\rm{p}}&=&\left\{\begin{array}{rl}
 7.70^{+11.34}_{-3.90} M_{\rm Jup} &\mbox{\rm{\ (close)}}\\
11.98^{+17.65}_{-6.31} M_{\rm Jup} &\mbox{\rm{\ (wide).}}
\end{array} \right.
\end{eqnarray}
The results indicate that the lens is a low-mass M dwarf with a super-Jupiter planet or a brown dwarf. The projected separation between the host and the planet is 
\begin{eqnarray}
a_{\perp}= sD_{\rm{L}} \theta_{\rm{E}}&=&\left\{\begin{array}{rl}
0.79^{+0.14}_{-0.16} \ \rm{au} &\mbox{\rm{\ (close)}}\\
1.50^{+0.27}_{-0.29} \ \rm{au} &\mbox{\rm{\ (wide).}} 
\end{array} \right.
\end{eqnarray}
Considering the snow line location of $a_{\rm snow}=2.7(M/M_\odot) \ \rm{au}$ (Kennedy $\&$ Kenyon\citealt{kennedy2008}), the planet orbits beyond the snow line of the host star in both models. The physical parameters of the lens system are described in Table 4.

\begin{deluxetable}{lcc}
\tablecolumns{3} \tablewidth{0pt} \tablecaption{\textsc{Lens parameters}}
\tablehead{\colhead{Parameters}     &   \colhead{close}   &  \colhead{wide} }
\startdata
$M_{\rm{h}} \ (M_{\odot})$       &  $0.09^{+0.14}_{-0.05}$       &  $0.09^{+0.14}_{-0.05}$      \\
$M_{\rm{p}} \ (M_{\rm{Jup}})$     &  $7.70^{+11.34}_{-3.90}$      &  $11.98^{+17.65}_{-6.31}$    \\
$D_{\rm{L}} \ (\rm{kpc})$         &  $7.01^{+1.05}_{-1.21}$       &  $7.04^{+1.05}_{-1.20}$      \\  
$a_{\perp} \ (\rm{au})$          &  $0.79^{+0.14}_{-0.16}$       &  $1.50^{+0.27}_{-0.29}$      \\ 
$\mu_{\rm{rel}} \ (\rm{mas} \ yr^{-1})$     &  $7.68 \pm 0.70$    &  $7.53 \pm 0.70$             \\ 
\enddata
\label{table:physical}
\end{deluxetable}

\bigskip
\section{Investigation of Bayesian Analysis}
There have now been several events with direct measurements of the lens flux from high-resolution imaging (Batista et al.\citealt{batista2015}, Bennett et al.\citealt{bennett2015}, Bhattacharya et al.\citealt{bhattacharya2019}, Bennett et al.\citealt{bennett2020}, Bhattacharya et al.\citealt{bhattacharya2020}, Terry et al.\citealt{terry2020}, Vandorou et al.\citealt{vandorou2020}). Comparing the inferred masses to the predictions from prior Bayesian estimates has often resulted in disagreements. At the same time, the true masses are not necessarily expected to agree perfectly because a Bayesian posterior produces a distribution. For example, Shan et al.\citet{shan2019} compared the measured masses to the Bayesian posteriors for a sample of events with masses measured from \textit{Spitzer} parallaxes. They found that although there were some nominal outliers relative to a Bayesian prediction, as an ensemble, they were consistent with being drawn from the distribution, e.g., $5\%$ of events should have masses at least $2\sigma$ from the central predicted value. At the same time, we would benefit from a better understanding of how the inputs to Bayesian analyses affect the outputs.

\begin{figure}
\plotone{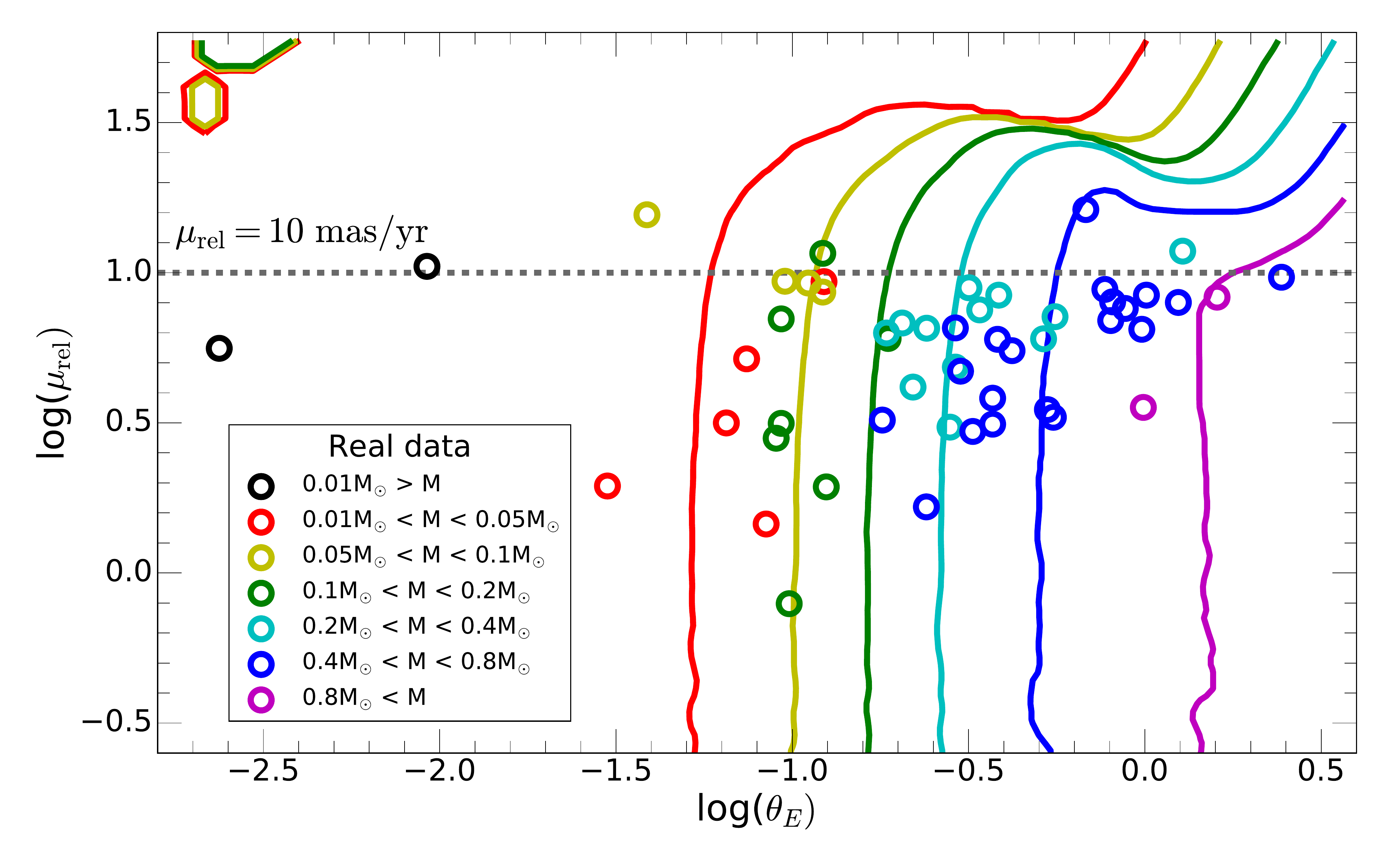}
\caption{Median mass produced by a Bayesian analysis for different combinations of $\theta_{\rm E}$ and $\mu_{\rm rel}$. The solid contours are shown at fixed masses $M=(0.03, 0.075, 0.15, 0.3, 0.6, 0.9)\ M_{\odot}$ (red, yellow, green, cyan, blue, magenta). The open circles are events from the literature with masses inferred from a Bayesian analysis; they are color-coded by mass. The contours indicate that for $\mu_{\rm rel} \lesssim 10\ \mathrm{mas\ yr}^{-1}$ ($\log \mu_{\rm rel} < 1$), the inferred mass is almost completely determined by $\theta_{\rm E}$.} 
\label{fig:thetaEmu}
\end{figure}

In order to investigate the roles of $\theta_{\rm E}$  and $\mu_{\rm rel}$ in the Bayesian analysis, we repeat the calculation of the predicted lens mass on a grid of $\log \theta_{\rm E}$ and $\log \mu_{\rm rel}$. Here, we adopt the ranges $-2.8 \leq \log \theta_{\rm E} \leq 0.6$ and $-0.8 \leq \log \mu_{\rm rel} \leq 1.8$, which span the range of values published in microlensing papers, and $(\log \theta_{\rm E}, \log \mu_{\rm rel})$ are uniformly divided over a $(100\times100)$ grid. Figure \ref{fig:thetaEmu} shows that for $\mu_{\rm rel} \lesssim 10\ \mathrm{mas\ yr}^{-1}$, the median mass derived from the Bayesian analysis depends only on $\theta_{\rm E}$. That is, once $\theta_{\rm E}$ is known, the inferred mass does not depend on the lens-source relative proper motion. 

At first, this seems surprising because it seems $\mu_{\rm rel}$ should constrain the Galactic population to which the lens belongs. If both the lens and the source are in the Galactic bulge, we might expect only a small relative proper motion between them. By contrast, if the lens is in the Disk, we might expect a larger relative proper motion on average because disk stars move with the galactic rotation curve relative to the bulge. If the lens is very nearby the Sun, we might expect an even larger proper motion due to the peculiar motion of the lens. However, a brief investigation reveals that only the third constraint (extremely large proper motions corresponding to very nearby lenses) is significant.

Generally, the source is in bulge. So, its mean proper motion is $<\mu_{\rm rel}(l,b)>\ = (-6,0)\ \mathrm{mas\ yr}^{-1}$ and the dispersion in each direction is about $3\ \mathrm{mas\ yr}^{-1}$. Thus, for bulge-bulge lensing, $<\mu_{\rm rel}(l,b)>\ = (0,0)\ \mathrm{mas\ yr}^{-1}$ as expected, but the dispersion in each direction is $4.2\ \mathrm{mas\ yr}^{-1}$. Hence, any $|\mu_{\rm rel}|$ is allowed in the range $0 < |\mu_{\rm rel}| < 9 \ \mathrm{mas\ yr}^{-1}$. For disk lenses, $<\mu_{\rm rel}(l,b)>\ =(6,0)\ \mathrm{mas\ yr}^{-1}$, but again, the dispersion in each direction is $3.3\ \mathrm{mas\ yr}^{-1}$ (note the dispersion of disk lenses is small and more-or-less
constant as a function of distance). So again, pretty much any
 $\mu_{\rm rel}$ is allowed for $0 < |\mu_{\rm rel}| < 10 \ \mathrm{mas\ yr}^{-1}$. Thus, $|\mu_{\rm rel}| \lesssim 10\ \mathrm{mas\ yr}^{-1}$ is not constraining, which leaves only the $\theta_{\rm E}$ constraint as contributing to the determination of the lens mass.
 
In terms of the Bayesian likelihood function, if we ignore $\mu_{\rm rel}$ (which we have established is not constraining), we have
\begin{equation}
\Gamma(M,D) = D^3 \rho(D) * g(M),
\end{equation}
where $\rho(D)$ is the density of stars and $g(M)$ is their mass function.
But, 
\begin{equation}
\theta_{\rm E}^{2} = \kappa \pi_{\rm rel} * M  \rightarrow M = \frac{\theta_{\rm E}^{2}}{\kappa\pi_{\rm rel}},
\end{equation}
where $\pi_{\rm rel} = \rm{au}/D - \pi_{\rm S}$ is the relative lens-source parallax. So, if $\theta_{\rm E}$ is known, then
\begin{equation}
\Gamma(M,D) \rightarrow \Gamma(D) =  D^{3} \rho(D)*g \left(\frac{\theta_{\rm E}^{2}}{\kappa \pi_{\rm rel}}\right),
\label{eqn:gammaD}
\end{equation}
where $\pi_{\rm rel}$ is regarded as an implicit
function of $D$, and $\pi_{\rm s}$ is regarded as fixed (or varying over
a narrow range (technically, Equation (\ref{eqn:gammaD}) should have another factor of $\theta_{\rm E}$ but this is always
the same because it is measured).
Hence, if we hold $\theta_{\rm E}$ fixed, then
$\Gamma(D)$ is some smooth function of $D$. One could take the peak of this function to be the most likely value, and hence, infer $M$. Alternatively, one could marginalize over $D$, in which case
\begin{equation}
D_{\rm best} = \frac{\Sigma_i D_i \Gamma(D_i)}{\Sigma_i \Gamma(D_i)}
\end{equation}
and
\begin{equation}
M_{\rm best} = \frac{\Sigma_i M(\theta_{\rm E}^2/\kappa \pi_{{\rm rel},i})\Gamma(D_i)}{\Sigma_i \Gamma(D_i)} .
\end{equation}
Either way, the resulting mass (and distance) depends solely on $\theta_{\rm E}$.

\begin{figure}
\includegraphics[width=0.49\textwidth]{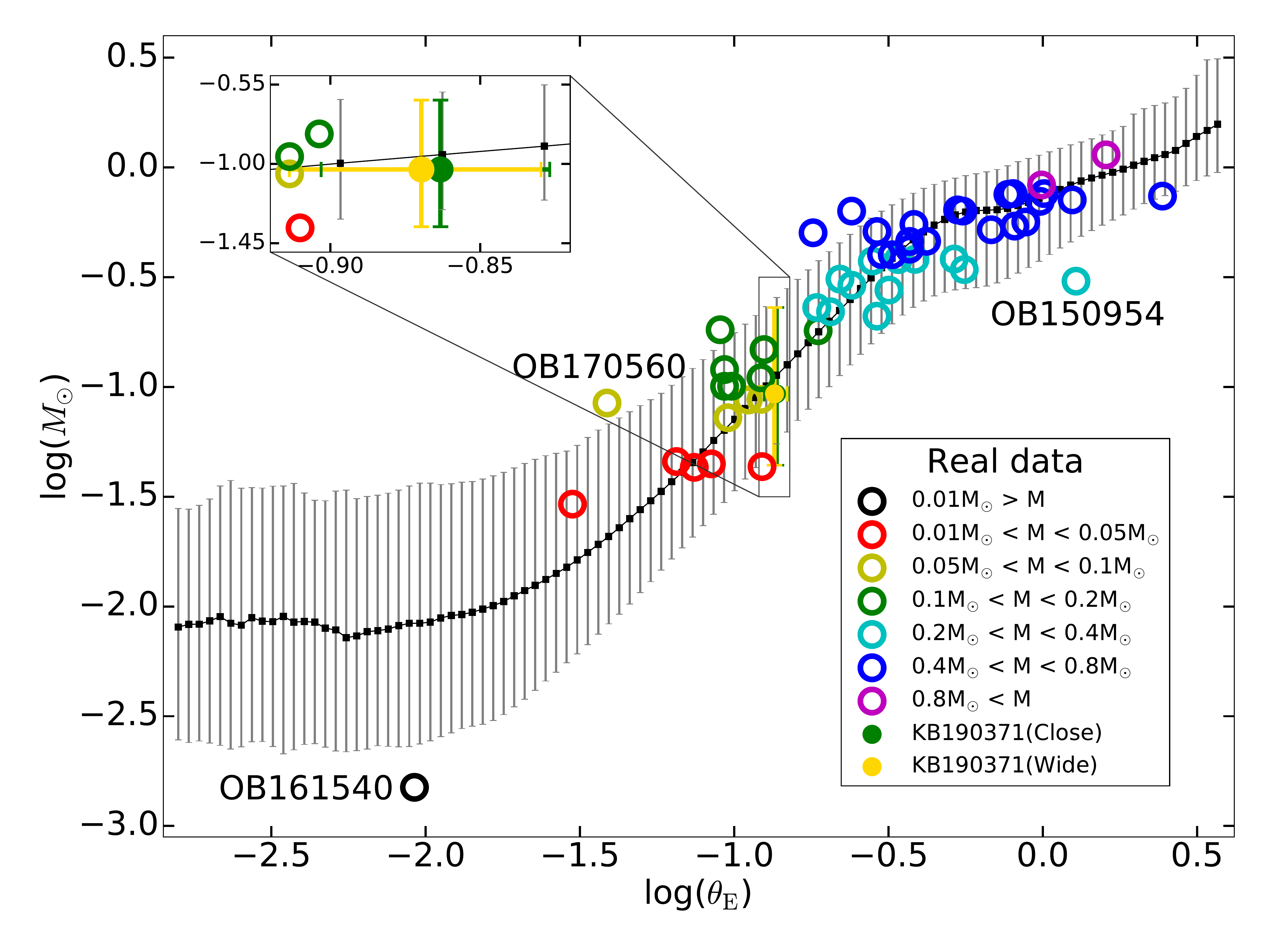}
\includegraphics[width=0.49\textwidth]{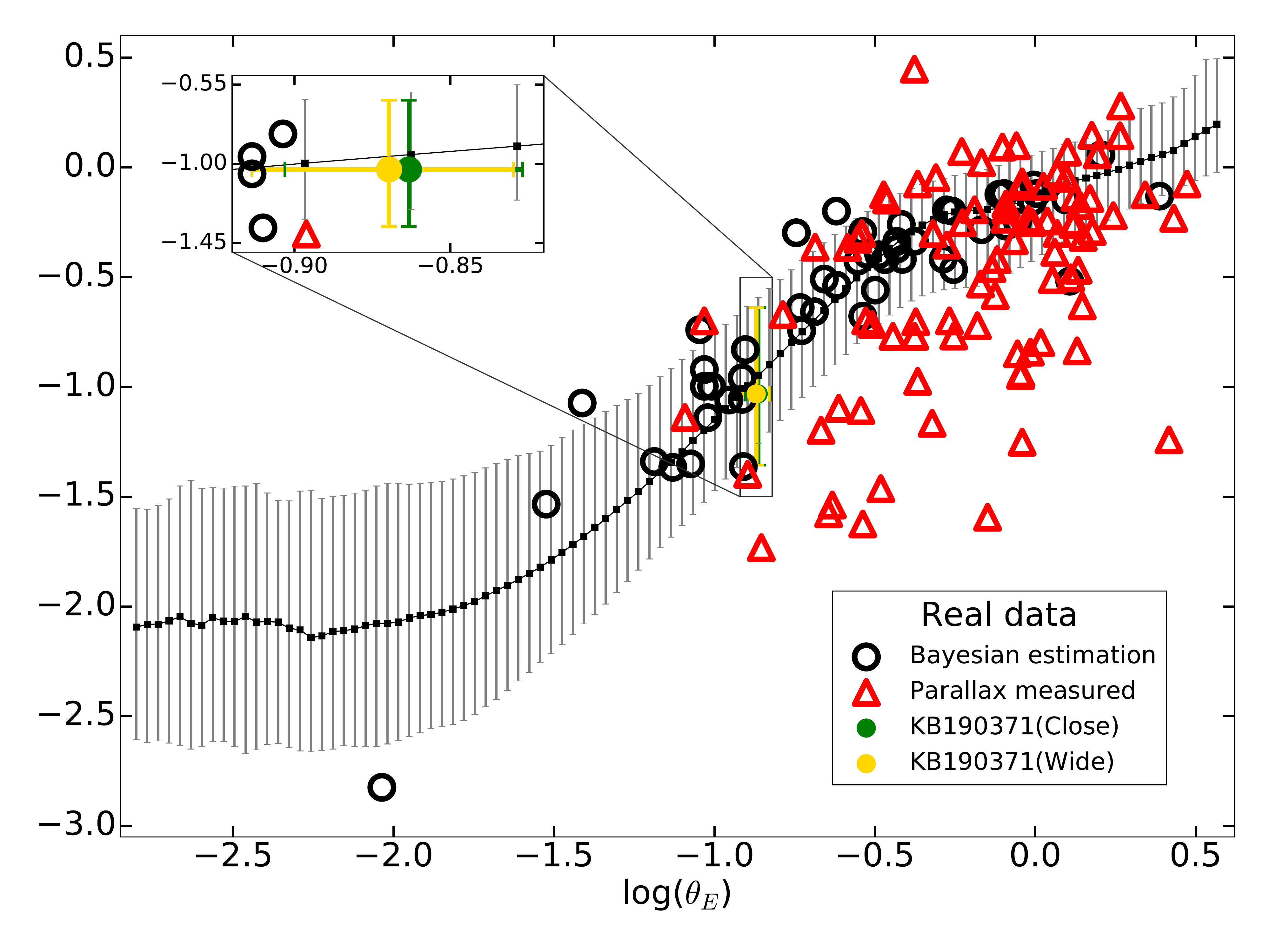}
\caption{Median Bayesian mass estimate as a function of $\theta_{\rm E}$, marginalized over $\mu_{\rm rel}$ for $\mu_{\rm rel} < 10\ \mathrm{mas\ yr}^{-1}$ (black points with errorbars). Open circles show that the Bayesian mass estimates from the literature are all consistent with these estimates despite being along different lines of sight and using a range of priors. The outliers all have $\mu_{\rm rel} > 10\ \mathrm{mas\ yr}^{-1}$, and so are not expected to obey this relation (see Figure \ref{fig:thetaEmu}). {\em Left}: Points are color-coded by mass. {\em Right}: lenses with mass estimates from a Bayesian analysis are plotted as black circles. Lenses that also have parallax information (so a mass measurement) are plotted as red triangles.}
\label{fig:thetaEmass}
\end{figure}

Figure \ref{fig:thetaEmass} shows the predicted lens mass (and its 68\% confidence interval) as a function of $\theta_{\rm E}$ after marginalizing over $\mu_{\rm rel}$ for $\mu_{\rm rel} < 10\  \mathrm{mas\ yr}^{-1}$. Although this relation was generated specifically for KMT-2019-BLG-0371 using the Galactic model described in Section 4.2, comparing to other published events with masses inferred from a Bayesian analysis shows that it is broadly applicable. That is, although the lens masses for the events from the literature\footnote{For events with multiple solutions, we used the mean mass for the different solutions.} were derived along different lines of sight and using different Galactic models, the scatter in the inferred masses is smaller than the intrinsic uncertainties in our relation. 

The three clear outliers on this plot (OB150954; Shin et al.\citealt{shin2016}, Bennett et al.\citealt{bennett2017}, OB161540; Mr\'{o}z et al.\citealt{mroz2018}, and OB170560; Mr\'{o}z et al.\citealt{mroz2019}) all have proper motions larger than our limit of $\mu_{\rm rel, max} = 10\ \mathrm{mas\ yr}^{-1}$. A fourth outlier (OB121323; Mr\'{o}z et al.\citealt{mroz2019}) is outside the bounds of this plot, but it has $\log\theta_{\rm E} = -2.63$ and an expected mass of $M\sim 2$--$20\ M_{\oplus}$. In this case, the Bayesian mass is highly sensitive to the choice of mass function, because the mass function for objects smaller than brown dwarfs is poorly constrained. In general, there are very few events with $\log \theta_{\rm E} < -1.0$, so it is unclear if $\theta_{\rm E}$ still dominates inferences about the lens mass, especially because this corresponds to the region where our knowledge of the mass function prior becomes highly uncertain. However, the lack of scatter for larger $\theta_{\rm E}$ implies that the relation shown in Figure \ref{fig:thetaEmass} provides almost as much insight into the nature of lens as a full Bayesian analysis based on measurements of $\theta_{\rm E}$ and $\mu_{\rm rel}$.

At the same time, Figure \ref{fig:thetaEmass} also calls into question whether or not the ``insight" gained from a Bayesian analysis is meaningful. The right panel of Figure \ref{fig:thetaEmass} shows that the distribution of objects with true mass measurements (from combining a measurement of $\theta_{\rm E}$ with a measurement of $\pi_{\rm E}$) is much broader than the Bayesian relation. In fact, the distribution of mass at a given $\theta_{\rm E}$ is fairly uniform, rather than concentrated toward the most likely Bayesian mass. Of course, lenses with small masses may be over-represented in this plot (since, at fixed $\theta_{\rm E}$, a smaller mass implies a larger, more easily measured, parallax), but it still emphasizes the value of true mass measurements for interpreting the nature of microlensing events.

\bigskip
\section{Summary}
We analyzed the microlensing event KMT-2019-BLG-0371 with data from three different surveys (OGLE, KMTNet, and MOA). The best-fit model light curve of the event, thanks to the high cadence observations, demonstrated an evident caustic-crossing around the peak in spite of the short event duration ($t_{\rm E} \sim 6.5$ days). Nevertheless, the event suffered from the $s \leftrightarrow 1/s$ degeneracy and the microlens parallax was not measured. Therefore, there is ambiguity in the determination of a concrete solution for the lens. A Bayesian analysis suggests that the host star is likely to have a mass on the border between low-mass stars and brown dwarfs, making the companion a super-Jupiter planet, despite the large mass ratio.

Motivated by investigating the physical interpretation of this event, we use it as a proxy to study the role of various inputs to the Bayesian analysis used to infer the lens mass (and distance). This analysis takes both $\mu_{\rm rel}$ and $\theta_{\rm E}$ as inputs. However, the analysis shown in Figure \ref{fig:thetaEmu} shows that only $\theta_{\rm E}$ influences the inferred mass for $\mu_{\rm rel} < 10\ \mathrm{mas\ yr}^{-1}$. Furthermore, comparing to Bayesian analyses in the literature shows that variations due to different lines of sight or different galactic models are smaller than the intrinsic uncertainty in the Bayesian analysis. Hence, for many applications, it is likely sufficient to use the relation and confidence interval in Figure \ref{fig:thetaEmass} rather than a full Bayesian analysis. All values presented in Figure \ref{fig:thetaEmass}, which are $\theta_{\rm E}$, mass, and $\pm 1\sigma$ of the mass, are listed in Tables $5\sim9$.

At the same time, events with measured masses (based on measurements of the microlens parallax), show a much broader distribution than indicated by this relation. This appears to conflict with the results of Shan et al.\citet{shan2019} who showed that the true distribution of masses for events with measured masses and \Spitzer\ parallaxes was consistent with a statistical sampling of the Bayesian mass posteriors derived for those events. One possible explanation is that \Spitzer\ parallax measurements are less biased toward low-mass events than measurements from annual parallax, which dominate the sample shown in Figure \ref{fig:thetaEmass}. It is also possible that the $68\%$ confidence interval does not fully reflect the shape of the posterior, which is broad and non-Gaussian (see, e.g., Figure \ref{fig:five}).

Adaptive optics observations that measure the flux from the host star and/or independently measure $\mu_{\rm rel}$ could be used to both resolve the nature of the lens in KMT-2019-BLG-0371 and further explore the limits of Bayesian analysis. For KMT-2019-BLG-0371, the lens has a $\sim 50\%$ probability of being a star. Thus, given that the source is faint, it is possible that excess light from the lens could be detected immediately to either estimate the flux from the lens or place an upper limit on it. However, if the lens is not resolved from the source, there is a risk of confusion if the excess light is in fact due to a companion to the lens (or source), rather than the lens itself (see, e.g., Bhattacharya et al.\citealt{bhattacharya2017}). This confusion can be mitigated by waiting until the lens and source separate.
In $K$-band with a 10-m telescope, KMT-2019-BLG-0371L should be resolved from the source by $\sim 2026$. With a 30-m telescope, its flux should be observable even for masses into the brown dwarf regime. If such measurements were made for a sizeable sample of events that also have Bayesian estimates for the lens mass (at present there are only a handful of such objects), this would serve as an independent test of the relation shown in Figure \ref{fig:thetaEmass}.

\bigskip
\section*{acknowledgement}
Work by Y. H. Kim was supported by the KASI (Korea Astronomy and Space Science Institute) grant 2021-1-830-08. This research has made use of the KMTNet system operated by the Korea Astronomy and Space Science Institute (KASI) and the data were obtained at three host sites of CTIO in Chile, SAAO in South Africa, and SSO in Australia. The OGLE project has received funding from the National Science Centre, Poland, grant MAESTRO 2014/14/A/ST9/00121 to A.U. The MOA project was supported by JSPS KAKENHI grant No. JSPS24253004, JSPS26247023, JSPS23340064, JSPS15H00781, JP17H02871, and JP16H06287. Work by C.H. was supported by the grants of National Research Foundation of Korea (2019R1A2C2085965 and 2020R1A4A2002885).

\bibstyle{apj.bst}


\begin{deluxetable}{ccccc}
\tablecolumns{4} \tablewidth{0pc} \tablecaption{\textsc{Parameters plotted in Figure \ref{fig:thetaEmass}}}
\tablehead{\colhead{$\log\theta_{\rm{E}}$}  &   \colhead{$\log M_{\rm{L}}$}   &  \colhead{$-1\sigma(\log M_{\rm{L}})$} & \colhead{$+1\sigma(\log M_{\rm{L}})$} & \colhead{References}} 
\startdata
-2.801 & -2.094 & 0.514 & 0.540 & - \\
-2.767 & -2.081 & 0.539 & 0.525 & - \\
-2.733 & -2.080 & 0.532 & 0.541 & - \\
-2.699 & -2.066 & 0.556 & 0.556 & - \\
-2.665 & -2.046 & 0.587 & 0.595 & - \\
-2.631 & -2.076 & 0.574 & 0.650 & - \\
-2.625 & -4.398 & - & - & $\rm{OB121323^{(1)}}$ \\
-2.597 & -2.085 & 0.555 & 0.624 & - \\
-2.563 & -2.051 & 0.566 & 0.592 & - \\
-2.529 & -2.067 & 0.548 & 0.607 & - \\
-2.495 & -2.069 & 0.570 & 0.617 & - \\
-2.461 & -2.045 & 0.626 & 0.594 & - \\
-2.427 & -2.071 & 0.582 & 0.631 & - \\
-2.393 & -2.069 & 0.560 & 0.586 & - \\
-2.359 & -2.072 & 0.553 & 0.555 & - \\
-2.325 & -2.099 & 0.542 & 0.581 & - \\
-2.291 & -2.107 & 0.552 & 0.633 & - \\
-2.257 & -2.142 & 0.520 & 0.672 & - \\
-2.223 & -2.134 & 0.522 & 0.625 & - \\
-2.189 & -2.115 & 0.535 & 0.616 & - \\
-2.155 & -2.110 & 0.524 & 0.617 & - \\
-2.121 & -2.103 & 0.534 & 0.619 & - \\
-2.087 & -2.088 & 0.551 & 0.618 & - \\
-2.053 & -2.077 & 0.561 & 0.625 & - \\
-2.036 & -2.824 & - & - & $\rm{OB161540^{(2)}}$ \\
-2.019 & -2.077 & 0.550 & 0.638 & - \\
-1.985 & -2.071 & 0.541 & 0.633 & - \\
-1.951 & -2.053 & 0.541 & 0.608 & - \\
-1.917 & -2.041 & 0.534 & 0.600 & - \\
-1.883 & -2.037 & 0.520 & 0.603 & - \\
-1.849 & -2.026 & 0.520 & 0.596 & - \\
-1.815 & -2.012 & 0.525 & 0.593 & - \\
-1.781 & -1.996 & 0.523 & 0.591 & - \\
-1.747 & -1.978 & 0.515 & 0.589 & - \\
-1.713 & -1.952 & 0.506 & 0.583 & - \\
-1.679 & -1.927 & 0.496 & 0.579 & - \\
-1.645 & -1.903 & 0.479 & 0.573 & - \\
-1.611 & -1.877 & 0.463 & 0.564 & - \\
\enddata
\tablerefs{(1) Mr\'{o}z et al.\citet{mroz2019} and (2) Mr\'{o}z et al.\citet{mroz2018}.
}
\label{table:thetaEmass1}
\end{deluxetable}

\begin{deluxetable}{ccccc}
\tablecolumns{4} \tablewidth{0pc} \tablecaption{\textsc{Parameters plotted in Figure \ref{fig:thetaEmass}}}
\tablehead{\colhead{$\log\theta_{\rm{E}}$}  &   \colhead{$\log M_{\rm{L}}$}   &  \colhead{$-1\sigma(\log M_{\rm{L}})$} & \colhead{$+1\sigma(\log M_{\rm{L}})$} & \colhead{References}} 
\startdata
-1.577 & -1.850 & 0.450 & 0.547 & - \\
-1.543 & -1.821 & 0.435 & 0.530 & - \\
-1.524 & -1.534 & - & - & $\rm{MB15337^{(3)}}$ \\
-1.509 &	  -1.788 & 0.428   & 0.522 & -    \\
-1.475 &	  -1.754 & 0.420   & 0.523 & -    \\
-1.441 &	  -1.718 & 0.408   & 0.521 & -    \\
-1.412 & -1.073 & - & - & $\rm{OB170560^{(4)}}$ \\
-1.407 &	  -1.681 & 0.398   & 0.513 & -    \\
-1.373 &	  -1.641 & 0.394   & 0.500 & -    \\
-1.339 &	  -1.600 & 0.389   & 0.487 & -    \\
-1.305 &	  -1.559 & 0.378   & 0.474 & -    \\
-1.271 &	  -1.518 & 0.369   & 0.461 & -    \\
-1.237 &	  -1.476 & 0.359   & 0.447 & -    \\
-1.203 &	  -1.432 & 0.352   & 0.439 & -    \\
-1.187 & -1.339 & - & - & $\rm{OB171522^{(5)}}$ \\
-1.169 &	  -1.388 & 0.345   & 0.435 & -    \\
-1.135 &	  -1.344 & 0.339   & 0.429 & -    \\
-1.130 & -1.366 & - & - & $\rm{MOA}$-$\rm{bin}$-$29^{(6)}$ \\
-1.101 &	  -1.296 & 0.336   & 0.420 & -    \\
-1.074 & -1.350 & - & - & $\rm{KB161107^{(7)}}$ \\
-1.067 &	  -1.244 & 0.336   & 0.410 & -    \\
-1.046 & -0.739 & - & - & $\rm{OB181011^{(8)}}$ \\
-1.033 &	  -1.196 & 0.330   & 0.401 & -    \\
-1.032 & -0.997 & - & - & $\rm{OB150051^{(9)}}$ \\
-1.032 & -0.921 & - & - & $\rm{OB180677^{(10)}}$ \\
-1.021 & -1.140 & - & - & $\rm{OB151771^{(11)}}$ \\
-1.009 & -0.997 & - & - & $\rm{OB161227^{(12)}}$ \\
-0.999 &	  -1.148 & 0.325   & 0.395 & -    \\
-0.965 &	  -1.098 & 0.321   & 0.384 & -    \\
-0.955 & -1.060 & - & - & $\rm{KB180748^{(13)}}$ \\
-0.931 &	  -1.048 & 0.320   & 0.373 & -    \\
-0.914 & -1.057 & - & - & $\rm{KB162142^{(14)}}$ \\
-0.914 & -0.958 & - & - & $\rm{MB11262^{(15)}}$ \\
-0.910 & -1.363 & - & - & $\rm{KB161820^{(16)}}$ \\
\enddata
\tablerefs{(3) Miyazaki et al.\citet{miyazaki2018}, (4) Mr\'{o}z et al.\citet{mroz2019}, (5) Jung et al.\citet{jung2018a}, (6) Kondo et al.\citet{kondo2019}, (7) Hwang et al.\citet{hwang2019}, (8) Han et al.\citet{han2019b}, (9) Han et al.\citet{han2016}, (10) Martin et al.\citet{martin2020}, (11) Zhang et al.\citet{zhang2020}, (12) Han et al.\citet{han2020d}, (13) Han et al.\citet{han2020a}, (14) Jung et al.\citet{jung2018b}, (15) Bennett et al.\citet{bennett2014}, and (16) Jung et al.\citet{jung2018b}.
}
\label{table:thetaEmass2}
\end{deluxetable}

\begin{deluxetable}{ccccc}
\tablecolumns{4} \tablewidth{0pc} \tablecaption{\textsc{Parameters plotted in Figure \ref{fig:thetaEmass}}}
\tablehead{\colhead{$\log\theta_{\rm{E}}$}  &   \colhead{$\log M_{\rm{L}}$}   &  \colhead{$-1\sigma(\log M_{\rm{L}})$} & \colhead{$+1\sigma(\log M_{\rm{L}})$} & \colhead{References}} 
\startdata
-0.904 & -0.830 & - & - & $\rm{MB11291^{(17)}}$ \\
-0.897 &	  -0.996 & 0.317   & 0.362 & -    \\
-0.863 &	  -0.947 & 0.312   & 0.355 & -    \\
-0.829 &	  -0.899 & 0.306   & 0.348 & -    \\
-0.795 &	  -0.849 & 0.303   & 0.339 & -    \\
-0.761 &	  -0.799 & 0.303   & 0.332 & -    \\
-0.745 & -0.298 & - & - & $\rm{KB162364^{(18)}}$ \\
-0.733 & -0.638 & - & - & $\rm{MB09411^{(19)}}$ \\
-0.727 & -0.744 & - & - & $\rm{MB10353^{(20)}}$ \\
-0.727 &	  -0.748 & 0.301   & 0.323 & -    \\
-0.693 &	  -0.700 & 0.298   & 0.317 & -    \\
-0.688 & -0.658 & - & - & $\rm{OB05390^{(21)}}$ \\
-0.659 &	  -0.651 & 0.297   & 0.308 & -    \\
-0.658 & -0.509 & - & - & $\rm{OB08210^{(22)}}$ \\
-0.625 &	  -0.602 & 0.298   & 0.296 & -    \\
-0.620 & -0.199 & - & - & $\rm{KB162397^{(23)}}$ \\
-0.619 & -0.537 & - & - & $\rm{OB120724^{(24)}}$ \\
-0.591 &	  -0.552 & 0.299   & 0.284 & -    \\
-0.557 &	  -0.504 & 0.300   & 0.272 & -    \\
-0.553 & -0.427 & - & - & $\rm{OB08355^{(25)}}$ \\
-0.538 & -0.678 & - & - & $\rm{MB08310^{(26)}}$ \\
-0.538 & -0.292 & - & - & $\rm{OB141760^{(27)}}$ \\
-0.523 &	  -0.457 & 0.300   & 0.258 & -    \\
-0.523 & -0.397 & - & - & $\rm{MB11322^{(28)}}$ \\
-0.499 & -0.559 & - & - & $\rm{OB170373^{(29)}}$ \\
-0.489 &	  -0.414 & 0.299   & 0.243 & -    \\
-0.488 & -0.398 & - & - & $\rm{OB120838^{(30)}}$ \\
-0.469 & -0.420 & - & - & $\rm{MB09319^{(31)}}$ \\
-0.454 &	  -0.371 & 0.301   & 0.228 & -    \\
-0.432 & -0.372 & - & - & $\rm{OB181700^{(32)}}$ \\
-0.432 & -0.337 & - & - & $\rm{OB08513^{(33)}}$ \\
-0.420 &	  -0.330 & 0.307   & 0.212 & -    \\
\enddata
\tablerefs{(17) Bennett et al.\citet{bennett2018}, (18) Han et al.\citet{han2020e}, (19) Bachelet et al.\citet{bachelet2012}, (20) Rattenbury et al.\citet{rattenbury2015}, (21) Beaulieu et al.\citet{beaulieu2006}, (22) Jeong et al.\citet{jeong2015}, (23) Han et al.\citet{han2020e}, (24) Hirao et al.\citet{hirao2016}, (25) Koshimoto et al.\citet{koshimoto2014}, (26) Bhattacharya et al.\citet{bhattacharya2017}, (27) Bhattacharya et al.\citet{bhattacharya2016}, (28) Shvartzvald et al.\citet{shvartzvald2014}, (29) Skowron et al.\citet{skowron2017}, (30) Poleski et al.\citet{poleski2020}, (31) Miyake et al.\citet{miyake2011}, (32) Han et al.\citet{han2020b}, and (33) Jeong et al.\citet{jeong2015}.
}
\label{table:thetaEmass3}
\end{deluxetable}

\begin{deluxetable}{ccccc}
\tablecolumns{4} \tablewidth{0pc} \tablecaption{\textsc{Parameters plotted in Figure \ref{fig:thetaEmass}}}
\tablehead{\colhead{$\log\theta_{\rm{E}}$}  &   \colhead{$\log M_{\rm{L}}$}   &  \colhead{$-1\sigma(\log M_{\rm{L}})$} & \colhead{$+1\sigma(\log M_{\rm{L}})$} & \colhead{References}} 
\startdata
-0.418 & -0.260 & - & - & $\rm{OB151670^{(34)}}$ \\
-0.415 & -0.419 & - & - & $\rm{KB191339^{(35)}}$ \\
-0.386 &	  -0.294 & 0.314   & 0.199 & -    \\
-0.377 & -0.337 & - & - & $\rm{OB131721^{(36)}}$ \\
-0.352 &	  -0.263 & 0.322   & 0.185 & -    \\
-0.318 &	  -0.237 & 0.332   & 0.175 & -    \\
-0.287 & -0.417 & - & - & $\rm{OB170173^{(37)}}$ \\
-0.284 &	  -0.217 & 0.341   & 0.167 & -    \\
-0.277 & -0.194 & - & - & $\rm{OB07368^{(38)}}$ \\
-0.260 & -0.199 & - & - & $\rm{OB03235^{(39)}}$ \\
-0.255 & -0.465 & - & - & $\rm{OB151649^{(40)}}$ \\
-0.250 &	  -0.203 & 0.349   & 0.165 & -    \\
-0.216 & -0.197 &  0.351 &  0.168 & -  \\
-0.182 & -0.195 &  0.346 &  0.175 & -  \\
-0.168 & -0.284 & - & - & $\rm{OB06238^{(41)}}$ \\
-0.148 & -0.193 &  0.334 &  0.183 & -  \\
-0.114 & -0.186 &  0.319 &  0.194 & -  \\
-0.114 & -0.123 & - & - & $\rm{MOA}$-$\rm{bin}$-$1^{(42)}$ \\
-0.097 & -0.119 & - & - & $\rm{KB170165^{(43)}}$ \\
-0.092 & -0.267 & - & - & $\rm{OB130132^{(44)}}$ \\
-0.080 & -0.174 &  0.307 &  0.200 & -  \\
-0.056 & -0.249 & - & - & $\rm{MB08379^{(45)}}$ \\ 
-0.046 & -0.159 &  0.297 &  0.200 & -  \\
-0.012 & -0.140 &  0.288 &  0.196 & -  \\
-0.009 & -0.155 & - & - & $\rm{OB170604^{(46)}}$ \\
-0.004 & -0.080 & - & - & $\rm{OB171375^{(47)}}$ \\
 0.004 & -0.119 & - & - & $\rm{KB190842^{(48)}}$ \\
 0.022 & -0.121 &  0.276 &  0.192 & -  \\
 0.056 & -0.101 &  0.266 &  0.188 & -  \\
 0.090 & -0.081 &  0.259 &  0.184 & -  \\
 0.095 & -0.148 & - & - & $\rm{OB160613^{(49)}}$ \\
 0.107 & -0.518 & - & - & $\rm{OB150954^{(50)}}$ \\
 \enddata
\tablerefs{(34) Ranc et al.\citet{ranc2019}, (35) Han et al.\citet{han2020c}, (36) Mr\'{o}z et al.\citet{mroz2017}, (37) Hwang et al.\citet{hwang2018}, (38) Sumi et al.\citet{sumi2010}, (39) Bennett et al.\citet{bennett2006}, (40) Nagakane et al.\citet{nagakane2019}, (41) Jeong et al.\citet{jeong2015}, (42) Bennett et al.\citet{bennett2012}, (43) Jung et al.\citet{jung2019}, (44) Mr\'{o}z et al.\citet{mroz2017}, (45) Suzuki et al.\citet{suzuki2014}, (46) Han et al.\citet{han2020e}, (47) Han et al.\citet{han2020e}, (48) Jung et al.\citet{jung2020b}, (49) Han et al.\citet{han2017}, and (50) Bennett et al.\citet{bennett2017}.
}
\label{table:thetaEmass4}
\end{deluxetable}

\begin{deluxetable}{ccccc}
\tablecolumns{4} \tablewidth{0pc} \tablecaption{\textsc{Parameters plotted in Figure \ref{fig:thetaEmass}}}
\tablehead{\colhead{$\log\theta_{\rm{E}}$}  &   \colhead{$\log M_{\rm{L}}$}   &  \colhead{$-1\sigma(\log M_{\rm{L}})$} & \colhead{$+1\sigma(\log M_{\rm{L}})$} & \colhead{References}} 
\startdata
 0.124 & -0.062 &  0.251 &  0.178 & -  \\
 0.158 & -0.048 &  0.239 &  0.179 & -  \\
 0.192 & -0.036 &  0.227 &  0.184 & -  \\
 0.205 &  0.057 & - & - & $\rm{OB181269^{(51)}}$ \\
 0.226 & -0.022 &  0.217 &  0.189 & -  \\
 0.260 & -0.008 &  0.209 &  0.195 & -  \\
 0.294 &  0.010 &  0.200 &  0.233 & -  \\
 0.328 &  0.028 &  0.192 &  0.240 & -  \\
 0.362 &  0.044 &  0.189 &  0.237 & -  \\
 0.387 & -0.131 & - & - & $\rm{OB110417^{(52)}}$ \\
 0.396 &  0.058 &  0.187 &  0.236 & -  \\
 0.430 &  0.078 &  0.186 &  0.243 & -  \\
 0.464 &  0.109 &  0.194 &  0.251 & -  \\
 0.498 &  0.141 &  0.201 &  0.279 & -  \\
 0.532 &  0.169 &  0.208 &  0.321 & -  \\
 0.566 &  0.196 &  0.220 &  0.298 & -  \\
\enddata
\tablerefs{(51) Jung et al.\citet{jung2020a} and (52) Shin et al.\citet{shin2012}.
}
\label{table:thetaEmass5}
\end{deluxetable}

\end{document}